\documentclass[usenatbib]{mn2e}
\usepackage{natbib}
\usepackage{epsf}
\usepackage{graphicx}
\newif\ifAMStwofonts

\title[Parallel tracks in X-ray binary outbursts]{Parallel tracks in infrared versus X-ray emission in black hole X-ray transient outbursts: a hysteresis effect?}
\author[Russell et al.] {David M. Russell, Thomas J. Maccarone, Elmar G. K\"ording,\\ School of Physics and Astronomy, University of Southampton, Highfield, Southampton, Hampshire, SO17 1BJ, UK,\\ davidr,tjm,elmar@astro.soton.ac.uk
\newauthor Jeroen Homan\\ Kavli Institute for Astrophysics and Space Research,\\ Massachusetts Institute
of Technology, 77 Massachusetts Avenue, 37-287, Cambridge, MA, USA, 02139, jeroen@space.mit.edu}

\begin{document}

\maketitle

\label{firstpage}

\def\simlt{\mathrel{\rlap{\lower 3pt\hbox{$\sim$}}
        \raise 2.0pt\hbox{$<$}}}
\def\simgt{\mathrel{\rlap{\lower 3pt\hbox{$\sim$}}
        \raise 2.0pt\hbox{$>$}}}

\input epsf

\begin{abstract}
We report the discovery of a new hysteresis effect in black hole X-ray binary
state transitions, that of the near-infrared (NIR) flux (which most likely originates in the jets) versus X-ray flux.  We find, looking at existing data sets, that the infrared emission of black hole X-ray
transients appears to be weaker in the low/hard state rise of an outburst than the low/hard state decline of an
outburst at a given X-ray luminosity. We discuss how this effect may be caused by a shift in the radiative efficiency of the inflowing or outflowing matter, or variations in the disc viscosity or the spectrum/power of the jet. In addition we show that there is a correlation (in slope but not in normalisation) between infrared and X-ray luminosities on the rise and decline, for all three low-mass black hole X-ray binaries with well-sampled infrared and X-ray coverage: $L_{\rm NIR}\propto L_{\rm X}^{0.5-0.7}$. In the high/soft state this slope is much shallower; $L_{\rm NIR}\propto L_{\rm X}^{0.1-0.2}$, and we find that the NIR emission in this state is most likely dominated by the viscously heated (as opposed to X-ray heated) accretion disc in all three sources.
\end{abstract}


\section{Introduction}

X-ray binary outbursts occur when there is a substantial increase of matter accreted from the outer accretion disc or companion star towards the compact object (a black hole or a neutron star). Spectral states in X-ray binaries were first discovered to exist in
the early 1970s (Tananbaum et al. 1972), when Cygnus X--1 showed a
sharp transition in its spectral energy distribution from a state
where its luminosity was dominated by soft $\gamma-$ray emission to one
where it was dominated by soft X-ray emission.  It was also found that
this transition corresponded to a turn-off in the observed radio
emission from Cygnus X-1 (Tananbaum et al. 1972), and it has since
been shown that this strong correlation between X-ray spectral shape
and strength of radio emission is ubiquitous (e.g. Fender et
al. 1999).  Furthermore, the shape of the X-ray spectrum of a source is
strongly correlated with its timing properties, with strong
variability generally correlated with strong radio and hard X-ray emission (see e.g. van der Klis 2006 for a review).

It was observed from early on that the different spectral states occur
at different fractions of the Eddington luminosity (see e.g. Nowak
1995 for a review). The low/hard state, which exhibits strong
radio emission, occurs at systematically lower luminosities than the
high/soft states with low amplitude variability and no radio emission.
Observations which fit neither of these states show similar
phenomenology over a rather wide range of luminosities; they have
typically been called intermediate states when at lower luminosities
and very high states when at high luminosities, but their spectral and
variability properties are sometimes comparable and sometimes varied \citep[see e.g.][]{homaet01,mcclet06,homabe05}.  More recently, a picture has begun to develop in which these spectral state transitions are hysteretic, with
the transition from hard state to soft state occurring at higher
luminosities than the transition from soft state to hard state
\citep*[see e.g.][]{miyaet95,nowaet02,barret02,smitet02,maccco03}.  At
sufficiently low luminosities, it appears that only low/hard states
can exist, but the brightest low/hard states are just as bright as the
brightest high/soft states (Homan \& Belloni 2005). At the same time, the spread in the
different soft-to-hard state transition levels for a variety of
sources appears to be considerably smaller than the spread in the
hard-to-soft state transition levels, with a clustering in
soft-to-hard state transition levels at about 2\% of the Eddington
limit (Maccarone 2003), although there do appear to be occasional
fainter transitions (e.g. Xue, Wu \& Cui 2006); one source, V4641 Sgr, shows behaviour that is not well fit by any of the canonical states at a luminosity of about $2\times 10^{-3}L_{\rm Edd}$ \citep{millet02}. For the most recent
effort at putting together a unified picture of spectral states in
X-ray binaries, we refer the reader to Fender, Belloni \& Gallo
(2004). More recently, it has been suggested that
the existence of a well-established relationship between X-ray
luminosity and radio luminosity in the low/hard state with $L_{\rm radio}
\propto L_{\rm X}^{0.7}$ (Corbel et al. 2000; Gallo, Fender \& Pooley 2003)
requires that the low/hard state be a radiatively inefficient flow,
with $L_{\rm X} \propto \dot{m}^2$, and with the radio emission providing a
good tracer of $\dot{m}$, the mass accretion rate \citep*{kordet06}. 

Both the rising low/hard state and the falling low/hard state connect
to a radiatively efficient flow in the soft state, where $L_{\rm X} \propto
\dot{m}$ \citep{shaksu73}, albeit at different luminosities. Changes in the
bolometric luminosity during the transitions between the hard and
soft state are generally less than 50 percent (see discussion in
\citealt{kordet06}, although a jump of more than 50 percent may
occur in some cases; \citealt{kaleet03}), suggesting that the
radiative efficiency of the accretion flow does not change sharply
during these transitions. The radius of the inner accretion disc may
also not change dramatically during transitions \citep[e.g.][]{fronet01a,millet06,rykoet07} although in most cases is likely to be truncated at low luminosities \citep*{mcclet05,esinet01,fronet01b,fronet03,chatet03,yuanet05,yuanet07}.

The X-ray luminosities in the rising
low/hard state and the falling low/hard state have the same
dependence on $\dot{m}$, but they also join to the same radiatively
efficient soft state (in which the X-ray luminosity has a different
$\dot{m}$ dependence). This could indicate different radiative
efficiency factors in the rising and falling low/hard state tracks.
One might then also expect parallel tracks in the jet power versus
X-ray luminosity in the two low/hard state tracks, related to such
efficiency differences. If some or all of the differences between the
two tracks are related to differences in $\alpha$, the dimensionless
viscosity parameter, then one might expect differences between the
tracks due to the theoretically expected $\alpha$ dependence in the
relation between jet power and mass accretion rate (see Meier 2001).
This would require $\alpha$ to change during the soft/intermediate
states between outburst rise and decline. Variations in $\alpha$
during transient outbursts are expected in the context of the disc
instability model for X-ray transient outbursts, and are thought to
be caused by a change in the ionisation state of hydrogen \citep*{meyeme81,cannet82,dubuet01}, or the dominant species in the case of
accretion discs in ultracompact X-ray binaries which are fed by
hydrogen poor white dwarfs (Menou, Perna \& Hernquist 2002).

One tentative report of parallel tracks in X-rays versus radio does
exist (Nowak et al. 2005), but in general, the inability of radio
telescopes to respond quickly to X-ray transients has led to most of
the radio observations of X-ray transients being made in the decaying
parts of the outbursts.  While this would, in principle, be the case
for optical telescopes as well, there have been dedicated monitoring
campaigns with small telescopes that have, in many cases, detected the
rises of outbursts from X-ray binaries in the optical and infrared
before they were detected in the X-rays, simply by monitoring known
X-ray transients in quiescence and waiting for them to go into outburst
(see e.g. Jain 2001; Jain et al. 2001).

There is mounting evidence for the spectrum of the jet to extend from the radio to infrared and even optical regimes in the low/hard state (and hence the rise and decay) from spectral \citep{hanet92,corbet01,chatet03} and timing \citep{kanbet01,uemuet04} studies. The near-infrared (NIR) emission has been found to be consistent with optically thin synchrotron emission \citep{homaet05,hyneet06} which appears to dominate the NIR regime (but probably not the optical) in this state \citep{corbet02,buxtet04,russet06} because the emission drops dramatically in the soft state, when the jet is expected to be quenched. We can therefore use the infrared flux to test whether there are parallel tracks in jet power versus X-ray power in X-ray binaries in the hard state.

\begin{figure}
\centering
\includegraphics[width=8.2cm,angle=0]{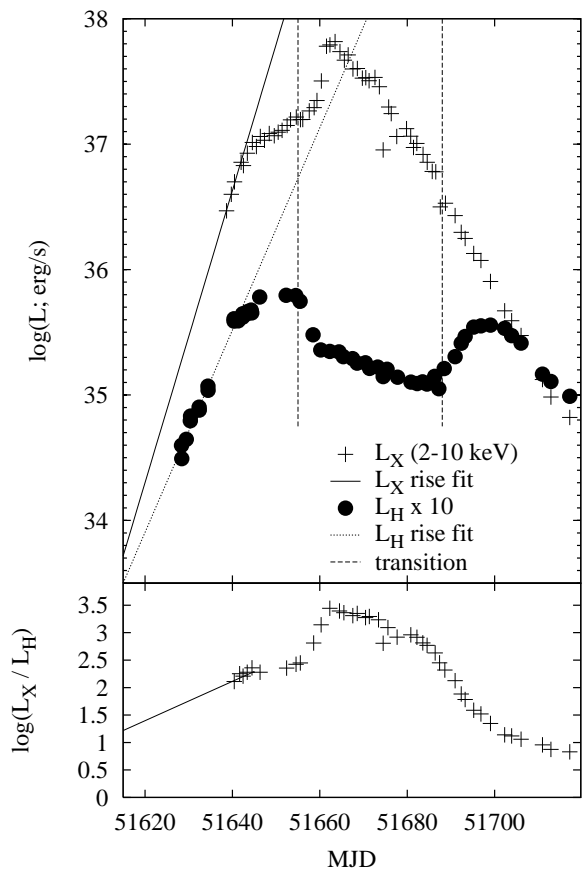}
\caption{The NIR and X-ray light curves of the 2000 outburst of XTE J1550--564 (upper panel). The lines are fit to the initial rise data before the break in the rise behaviour of both light curves. All X-ray data are from the \emph{RXTE} ASM except in the hard state decline, which are from the \emph{RXTE} PCA/HEXTE. The ratio of X-ray to NIR fluxes are shown in the lower panel.}
\end{figure}

\begin{table*}
\begin{center}
\caption{The sources used and data collected to calculate their luminosities.}
\begin{tabular}{lllllll}
\hline
Source &Outburst&Telescopes&Distance&$A_{\rm V}$&$N_{\rm H}$ /    &Flux\\
       &        &          &/ kpc   &           &10$^{21} cm^{-2}$&refs\\
\hline
\vspace{1mm}
4U 1543--47 &2002; decline&YALO 1.0 m, \emph{RXTE} PCA/HEXTE/ASM&7.5$\pm$0.5 &1.55$\pm$0.15&4.3$\pm$0.2&1, 2\\
\vspace{1mm}
XTE J1550--564 &2000; rise + decline&YALO 1.0 m, \emph{RXTE} PCA/HEXTE/ASM&5.3$\pm$2.3&5.0&8.7$\pm$2.1&3, 4\\
\vspace{1mm}
GX 339--4 &2002; rise&YALO 1.0 m, \emph{RXTE} PCA/ASM&8$^{+7.0}_{-1.0}$&3.9$\pm$0.5&6$^{+0.9}_{-1.7}$&5\\
GX 339--4 &1999; decline&ATCA$^1$, \emph{RXTE} PCA& & & &6\\
\hline
\end{tabular}
\normalsize
\end{center}
$^1$Australia Telescope Compact Array. The distances and extinction estimates are from the references in Table 1 of \cite{russet06}, except $A_{\rm V}=5$ for XTE J1550--564 \citep[][ see text]{tomset01,tomset03,kaaret03}. References for the fluxes used:
(1) \cite*{buxtet04};
(2) \cite*{kaleet05};
(3) \cite*{jainet01};
(4) \cite*{millet01};
(5) \cite*{homaet05};
(6) \cite*{nowaet05}
\end{table*}

\section{The data}

We searched the literature for BHXBs with well sampled quasi-simultaneous ($\Delta t \leq 1$ day) NIR--X-ray or radio--X-ray data on the rise and/or decline of a transient outburst (where NIR refers to $J$, $H$ or $K$-band). Data from three sources were found: 4U 1543--47, XTE J1550--564 and GX 339--4. XTE J1550--564 is the only source for which we obtain both rise and decay data of the same outburst. Apparent fluxes were converted to intrinsic luminosities adopting the best known estimates of the distance to the sources, the extinction $A_{\rm V}$ and neutral hydrogen column density $N_{\rm H}$. In Table 1 we list these values and the outbursts, telescopes used and references of the data. We adopt $A_{\rm V}=5$ for XTE J1550--564 as opposed to $A_{\rm V}=2.5$, which is adopted in \cite{russet06}. A value of $A_{\rm V}\sim 5$ is inferred from the hydrogen column density measured from X-ray observations of this source \citep*{tomset01,tomset03,kaaret03}. A value of $A_{\rm V}\sim 2.5$ produces a red optical spectrum \citep[$\gamma < 0$ where $F_{\nu}\propto \nu^{\gamma}$;][]{russet06}, which would indicate a much cooler disc than is expected, whereas with $A_{\rm V}\sim 5$ the spectrum is blue ($\gamma > 0$), as is expected for the Rayleigh-Jeans side of the disc blackbody spectrum. While uncertainties in the extinction will affect the relative normalisations of the $L_{\rm NIR}$--$L_{\rm X}$ relations for different sources, they will not affect the slopes of the relations for the individual sources.

To calculate the X-ray unabsorbed 2--10 keV luminosity we assume a power-law with a spectral index $=-$0.6 (photon index $\Gamma=1.6$) when the source is in the hard state and a blackbody at a temperature of 1 keV when the source is in the soft state \citep[these are the same as in][]{gallet03,russet06}. The resulting luminosities do not change significantly if similar reasonable approximations to these values are adopted. \emph{RXTE} ASM X-ray counts were converted to flux units using the \emph{NASA} tool \emph{Web-PIMMS}. Since we use data close to state transitions, we adopt the dates of these transitions as defined by the analysis of the authors in the literature. Inaccurate background subtraction could severely affect the X-ray luminosity estimates at low luminosities. For 4U 1543--47, \cite{kaleet05} have accounted for the Galactic X-ray background. The X-ray data we use for XTE J1550--564 is more than two orders of magnitude above the quiescent level so no background subtraction is necessary. The X-ray background contamination for GX 339--4 is uncertain \citep{nowaet02}, but for a reasonable choice of background fluxes the slope of the resulting NIR--X-ray correlation (see Section 3) does not change.

For NIR data near quiescence (or close to the lowest NIR fluxes observed from the source) we only include data which are $\geq 0.5$ mag above the measured quiescent flux level. We also subtract this quiescent level from all NIR data so that only the outburst flux remains (the NIR in quiescence is likely dominated by the companion star). The well-sampled NIR quiescent level of XTE J1550--564 is brighter by $\sim 0.25$ magnitudes after the 2000 outburst than prior to it \citep[Fig 1a of][]{jainet01}. This enhancement could be due to thermal emission from the heated accretion disc, or jet emission.

It is possible that the star or outer regions of the accretion disc are re-emitting X-ray radiation which was absorbed during the outburst.  The donor star in XTE J1550--564 \citep[which has spectral type G8 to K4;][]{oroset02} occupies around 10 percent of the solid angle seen by the X-ray source and should absorb $\sim 70$ percent of the radiation incident on it \citep{basksu73}.  Since the outburst gave off $\sim 3\times10^{43}$ ergs and the source brightened by about $10^{33}$ ergs s$^{-1}$ in the infrared, it could maintain this brightening for $\sim 70$ years if all the energy was lost in this infrared waveband. The brightening was present in at least the optical $V$-band too, but the brightening of the broadband optical--infrared spectrum could still be sustained for more than one year, longer than the duration of the quiescent light curve used in this paper. In addition, \cite{oroset02} found that the quiescent optical flux level of XTE J1550--564 was lower in May 2001 than before \emph{and} after the 2000 outburst. We therefore have subtracted the (not-dereddened) base levels of $H=16.25$ from the data on the hard state rise and $H=16.0$ from the hard state decline data. We also note that long-term variations in infrared quiescent levels have been observed in GRO J0422+32, A0620--00 and Aql X--1 (Reynolds, Callanan \& Filippenko 2007; C. Bailyn, private communication).

NIR intrinsic \citep*[de-reddened using the extinction $A_{\rm V}$ listed in Table 1, according to][]{cardet89} luminosities were calculated adopting the approximation $L_{\rm NIR}\approx \nu F_{\nu,NIR}$ \citep[we are approximating the spectral range of each filter to the central wavelength of its waveband; the same method as][]{russet06}.

\section{Results}

The X-ray and NIR light curves of the 2000 outburst of XTE J1550--564 are plotted in Fig. 1 \citep[upper panel; see also][]{jainet01}. In both X-ray and NIR light curves there is a break in the relation between luminosity and time on the outburst rise. One explanation for this could be that the point at which the luminosity versus time curve flattens corresponds to the start of the hard/intermediate state, and not to the bona fide low/hard state.  However, this traditionally occurs when the X-ray spectrum begins to soften, which happened about six days after the X-ray break (the first \emph{transition} dashed line in Fig. 1; see also \citealt{millet01}). We note also that the flattening in the infrared appears to happen a few
days before the flattening in the X-rays, and the optical ($V$-band) precedes the NIR ($H$-band) also by $\sim 3$ days \citep{jainet01}.  One may expect these delays to reveal a relation between wavelength and delay time, but the order between the wavebands is incorrect for this to be the case; the optical change is followed by the infrared and then X-ray. The reasons for these delays between wavebands are not clear. \cite{uemuet00} found that the X-rays lagged the optical rise of an outburst of XTE J1118+480 by about 10 days, and interpreted it as in `outside-in' outburst; the optical emission originating from the viscously-heated (as opposed to X-ray heated) accretion disc. In the case here of XTE J1550--564, we would not expect the NIR light to lag the optical, as is observed, if the viscous disc is the origin of the emission.

We have fit exponentials in luminosity versus time for the initial rise data before the break to a shallower slope in both light curves (Fig. 1 upper panel). These are required to obtain the $L_{\rm NIR}$--$L_{\rm X}$ relation before the break, as there is little quasi-simultaneous data on the rise but clear luminosity--time relations exist. For the X-ray light curve we have made a fit to the first four \emph{RXTE} ASM data points (all have $>3\sigma$ significance). The break appears just after these data in the X-rays and all four points lie very close to the fit. All $H$-band data before the gap at MJD 51634--40 were used for the $L_{\rm NIR}$--time fit.

\begin{figure}
\centering
\includegraphics[width=8.8cm]{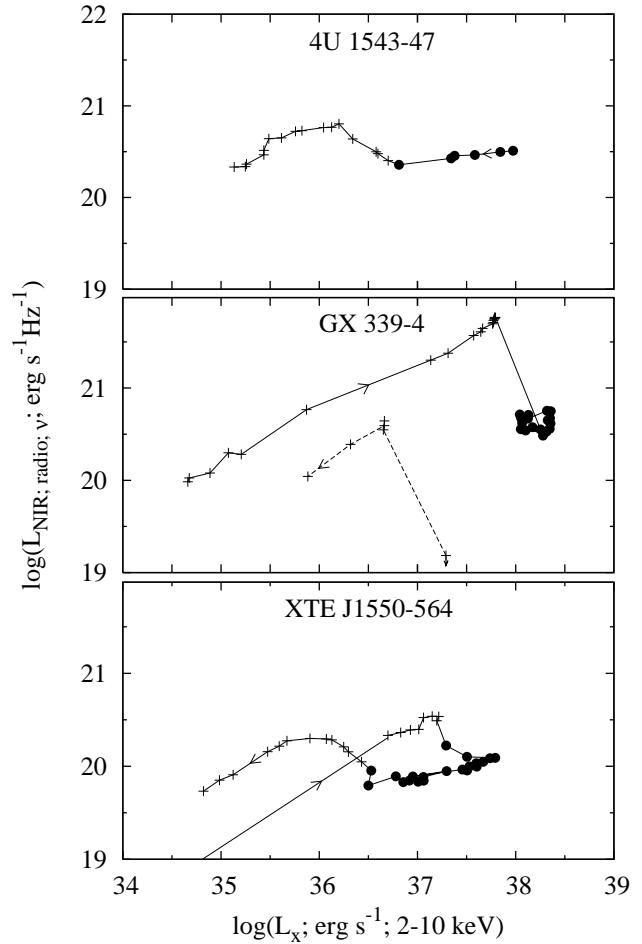}
\caption{X-ray luminosity versus quasi-simultaneous monochromatic NIR/radio luminosity for the three sources. Crosses are hard state data and filled circles represent data in the soft state when the jet is quenched. The NIR data are $J$-band for 4U 1543--47 and $H$-band for XTE J1550--564 and GX 339--4. The data on the decline for GX 339--4 are radio 8.64 GHz from a different outburst, not NIR (and is the only radio data we use).}
\end{figure}

\begin{figure}
\centering
\includegraphics[height=8.2cm,angle=270]{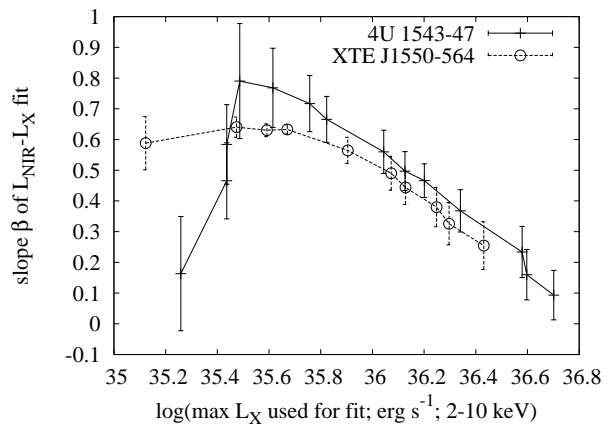}
\caption{The slope $\beta$ of the $L_{\rm NIR}$--$L_{\rm X}$ relation for the decline data of 4U 1543--47 and XTE J1550--564, as a function of the maximum $L_{\rm X}$ of the data used to infer the fit.}
\end{figure}

In Fig. 2 we plot the quasi-simultaneous NIR (or radio) and X-ray luminosities for each source. The luminosity--time fits to the rise of the XTE J1550--564 outburst in NIR and X-ray (Fig. 1) are used to infer the $L_{\rm NIR}$--$L_{\rm X}$ relation on the rise for this source before the data are sampled well. This fit has a slope $L_{\rm NIR}\propto L_{\rm X}^{\beta}$ where $\beta=0.7$. Interestingly, fitting the data after the aforementioned NIR and X-ray breaks (and before the state transition) in the XTE J1550--564 light curve yields the same slope: $\beta=0.7$.

For GX 339--4 we have NIR rise data and radio decline data from different outbursts \citep{homaet05,nowaet05}. We therefore cannot compare the normalisations of the rise and decline in this source even if we assume the two outbursts were similar hysteretically; assumptions are required of the radio-to-NIR spectral index. For 4U 1543--47 we have outburst decline data only, but there is an increase in the NIR flux when the source enters the hard state, as is expected from jet emission \citep{homaet05,russet06}.

\begin{table}
\begin{center}
\caption{Measured slopes of the $L_{\rm NIR}$--$L_{\rm X}$ (or $L_{\rm radio}$--$L_{\rm X}$) relations. $L_{\rm NIR}\propto L_{\rm X}^{\beta}$.}
\begin{tabular}{lll}
\hline
Source &Data&Slope $\beta$\\
\hline
4U 1543--47&soft state&0.13$\pm$0.01\\
4U 1543--47&hard state decline&0.5--0.8\\
GX 339--4 &hard state rise&0.55$\pm$0.01\\
GX 339--4 &hard state decline$^2$&0.70$\pm$0.06\\
XTE J1550--564&hard state rise&0.69$\pm$0.05\\
XTE J1550--564&soft state$^1$&0.23$\pm$0.02\\
XTE J1550--564&hard state decline&0.63$\pm$0.02\\
\hline
\end{tabular}
\normalsize
\end{center}
$^1$All soft state data are used for this fit except those above the apparent relation at the beginning and end of the soft state phase (likely to be during state transition). $^2$The slope here is measured from the five radio data points.
\end{table}

We see from Fig. 2 that the NIR is enhanced at a given X-ray luminosity on the decline compared to the rise for the only source with both rise and decline data from the same outburst, XTE J1550--564 (based on the extrapolation of the X-ray data on the rise, as explained above). The level of enhancement is $\sim 0.7$ dex in NIR luminosity, or a factor of $\sim 5$. The NIR luminosity at a given $L_{\rm X}$ (i.e. the normalisation) also differs between sources; this may reflect distance and/or interstellar absorption uncertainties.

A striking observation from Fig. 2 is the similarity of the slope of the $L_{\rm NIR}$--$L_{\rm X}$ (or $L_{\rm radio}$--$L_{\rm X}$) relation in the low/hard state rise and decline between sources. In Table 2 we tabulate the slopes measured for each source. The slope of the relation for 4U 1543--47 and XTE J1550--564 on their declines depends on which hard state data are used, since there is an initial NIR rise after the transition to the hard state. In Fig. 3 we show the inferred slope $\beta$ as a function of the range of data used to measure the fit. We use all the data at low X-ray luminosity and impose a cut at a higher X-ray luminosity: $L_{\rm X}^{\rm max}$. We plot $\beta$ versus log($L_{\rm X}^{\rm max}$) in Fig. 3. There appears to be a smooth relation between $\beta$ and $L_{\rm X}^{\rm max}$; the slope becomes shallower when we include the data at high $L_{\rm X}$ because of the initial NIR rise as the source declines in $L_{\rm X}$. The cut required to measure the $L_{\rm NIR}$--$L_{\rm X}$ relation once the two wavebands are positively correlated should be at a value of $L_{\rm X}$ after the NIR begins to drop. At this cut we have $\beta \sim 0.5$--0.8 and $\beta = 0.63\pm0.02$ for 4U 1543--47 and XTE J1550--564, respectively. From Table 2 we can see that generally, the slopes in the hard state ($\beta = 0.5$--0.7) are in agreement with the radio--X-ray and optical/NIR--X-ray correlations found from these and other sources \citep{corbet03,gallet03,gallet06,russet06}.

\begin{figure}
\centering
\includegraphics[width=8cm,angle=0]{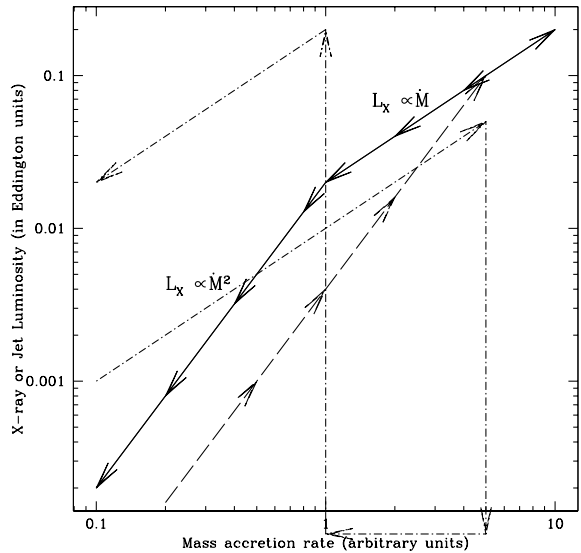}
\caption{A schematic diagram of the X-ray luminosity versus accretion
rate.  The units of the accretion rate are such that, in the
`normal' low/hard state, at the state transition \emph{back to} the low/hard state, the X-ray
luminosity is 2\% of the Eddington luminosity and the mass accretion
rate is unity.  The solid line corresponds to the decaying low/hard
state, and to sources like Cygnus X-1 which show no state transition/luminosity hysteresis
effects \citep{smitet02}, while the dashed line corresponds to the rising portion of
hysteretic light curves.  The arrows trace a typical outburst's track
in this diagram.  The dot-dashed line shows the evolution of the jet
kinetic power, as inferred from the observed hysteresis effect.  It drops below the $y$-axis in the high soft state to
indicate that the jet is `quenched' in this state.}
\end{figure}

In addition, Fig. 2 shows that all the NIR and radio data we use are suppressed in the soft state, indicating its most likely origin in the low/hard state is the jet, which is expected to be quenched (see Section 1). In addition, the optical--NIR colours are red ($\gamma < 0$) in all three sources in the hard state, indicating optically thin emission \citep{jainet01,buxtet04,homaet05}. The optical and NIR luminosity from the viscously-heated accretion disc should vary relatively smoothly across the state transition, as should X-ray reprocessing in the disc unless it largely depends on the hard spectral component (see Russell et al. 2006 for a more detailed discussion of these and other sources). The emission may also not be smooth if it is correlated with $\alpha$ and if $\alpha$ changes across the transition (see Section 4.2). Since we see a stronger soft state quenching in the NIR than in the optical wavebands in our sources \citep{jainet01,buxtet04,homaet05}, the longer wavelengths have much larger ratios of jet to disc contributions (this is expected since the spectral index of the jet spectrum is less than that of the disc).

In Table 2 we also state the slopes of the $L_{\rm NIR}$--$L_{\rm X}$ relations in the soft state for 4U 1543--47 and XTE J1550--564. The theoretically expected slope if the origin of the NIR emission is X-ray reprocessing in the disc is $\beta = 0.5$ \citep{vanpet94}. The shallower slopes observed are close to those expected if the origin of the NIR emission is the viscously heated disc: $\beta = 0.15$ \citep{russet06}. Since we are using a finite X-ray energy range (2--10 keV) as opposed to the bolometric X-ray luminosity, $\beta$ may be inaccurate because the fraction of the bolometric X-ray luminosity in this range will change with $L_{\rm X}$ in the soft state. However, this fraction would have to change by a factor of several during the $\sim$ one dex drop in $L_{\rm X}$ for $\beta$ to be inaccurate by a factor of $\sim 3$--4 (as is required if $\beta \sim 0.5$ for X-ray reprocessing). We can therefore conclude from correlation analysis that the viscously heated disc is likely to dominate the NIR in the soft state in these two sources. \cite{homaet05} also found this to be the dominating NIR emission process for GX 339--4 in the soft state (from the same data that we use here) since changes in the NIR preceded those in the X-ray by several weeks.

It is unlikely that disc emission contaminating the infrared luminosity plays a substantial role in the observed hysteresis effect in XTE J1550--564 in the hard state, because the separation between the rise and decline parallel tracks is a factor of about five in $L_{\rm NIR}$; the disc component would have to be dominating the NIR luminosity in the rising hard state for this to be the explanation of the effect. It would then be quite surprising for the colours to change so dramatically at the state transition and would have to be for some other reason than the quenching of the jet. In the decaying
hard state of XTE J1550--564, a rise is seen in $I$ and $V$ as well as in $H$ at the
time of the state transition, indicating again that the jet
power-to-disc power ratio is larger for the decaying hard state than
for the rising hard state (see the discussion of the optical
`reflare' in Jain et al. 2001).

\section{Discussion}

In this Section we concentrate mainly on interpreting the hysteresis effect. Nearly all models for jet production suggest that the kinetic power
supplied to the relativistic jet will be linearly proportional to the
mass accretion rate (e.g. Falcke \& Biermann 1995; Meier 2001). The observed monochromatic luminosity of the synchrotron self-absorbed part of the jet spectrum will be proportional
to the kinetic luminosity of the jet to the 1.4 power (e.g. Blandford \&
K\"onigl 1979; Falcke \& Biermann 1995; Heinz \& Sunyaev 2003).  As a
result, infrared and longer wavelength fluxes may be expected to
be excellent tracers of the mass accretion rate, and would be much
less sensitive to changes in the accretion efficiency than at X-ray
wavelengths \citep{kordet06}. However, in at least one neutron star X-ray binary 4U 0614+09, the NIR jet contribution is optically thin with the break between optically thick and optically thin emission lying further into the mid-infrared \citep{miglet06}. For BHXBs, it has been shown that this break lies close to the NIR \citep{corbet02,russet06} but may change with luminosity \citep{nowaet05}.

For a jet being powered from an advection
dominated accretion flow (ADAF; Narayan \& Yi 1995), it has been suggested
that the relation between $\dot{m}$ and the jet's kinetic luminosity
can also depend on the dimensionless viscosity parameter $\alpha$, and
on some fudge factors related to the fact that the standard ADAF treatment of
rotational velocities and azimuthal magnetic field strengths are only
approximations (Meier 2001).  This assumes that an ADAF produces the X-ray emission. Narayan
\& Yi (1995) did not account for the fact that shear in the Kerr
metric will enhance azimuthal magnetic fields, even for black holes
rotating at a fraction of their maximal rate (e.g. Meier 1999). We then find that:
\begin{equation}
P_{\rm J} \propto \dot{m}\alpha^{-1}f(g,j), 
\end{equation}
where $P_{\rm J}$ is the total power injected into the jet, $\dot{m}$ is the
mass accretion rate through the disc, and $f(g,j)$ is a correction
term which is a function of the black hole spin $j$ and the fudge factors which are rolled into $g$.

In the following subsections we use these arguments and discuss possible explanations for the observed hysteresis effect.

\subsection{Hysteresis traces the radiative efficiency of the accretion flow?}

Let us first consider the simplest possible interpretation of the
parallel tracks in XTE J1550--564 -- that they represent variations in the
radiative efficiency in the rising and decaying low/hard states, with
the mass accretion rate being traced perfectly by the infrared flux.
Here, we effectively assume that the variations in $\alpha$ and the
fudge factors are unimportant, as are variations in the radiative
efficiency of the jet itself for producing infrared emission.  Since
both the rising low/hard state and the decaying low/hard state are
observed to follow the $L_{\rm radio} \propto L_{\rm X}^{\sim 0.7}$ relation (Gallo et
al. 2003; 2006), they are both expected to have $L_{\rm X} \propto \dot{m}^2$ (a radiatively inefficient flow).  At
the same time, they are both expected to join to the same high/soft
state solution, which is believed to be a radiatively efficient ($L_{\rm X} \propto \dot{m}$)
geometrically thin accretion disc (Shakura \& Sunyaev 1973).  XTE J1550--564 was said to be in an intermediate state between the rise and decline of the 2000 outburst based on timing properties and a power law component in the X-ray spectrum. However, the blackbody flux was always dominant over the power law flux \citep{millet03}, and the radio/infrared jet was quenched so for purposes of this work we can treat it as a high/soft state from its jet properties \citep[see e.g.][]{fendet04} in the context of our following analysis. A
schematic version of the expected track followed by a hysteretic
source is plotted in Fig. 4.  We then see that the scenario where
$\dot{m}$ is well traced in the hard state by the infrared luminosity
and in the soft state by the X-ray luminosity is immediately
invalidated as this would require the infrared luminosity to be lower at a given $L_{\rm X}$ on the decline compared to the rise -- the opposite to what is observed. To achieve what is observed, we instead require the jet (NIR) luminosity to evolve as the dot-dashed lines in Fig. 4.

\subsection{Hysteresis traces $\alpha$?}

Next, let us consider the possibility that $\dot{m}$ is instead sensitive to variations in $\alpha$ ($P_{\rm J}$ and $\alpha$ are linked through equation 1).  This is a natural explanation for the different
state transition luminosities in the rising and falling low/hard
states; it is found by Esin, McClintock \& Narayan (1997), for
example, that the state transition luminosity should be 1.3$\alpha^2$,
while it is found by Zdziarski (1998) that the state transition
luminosity should scale approximately as $y^{3/5}\alpha^{7/5}$, where
$y$ is the Kompaneets parameter, and is defined as $y=4kT/(m_{\rm e} c^2){\rm max}(\tau,\tau^2),$
where $m_{\rm e}$ is the electron rest mass, $k$ is the Boltzmann constant,
$T$ is the temperature of the Comptonising region, and $\tau$ is the
optical depth of the Comptonising region. In either of these cases, then, one
would qualitatively expect that increasing $\alpha$ would increase the
state transition luminosity.  A similar result, albeit without an
analytic approximation for the size of the state transition
luminosity, is found by Meyer-Hofmeister, Liu \& Meyer (2005), who
showed that hysteresis may result from the different efficiencies of
hard photons and soft photons in evaporating mass from the thin disc
into the corona.

Some of the other hysteresis properties of outbursts seem to be
qualitatively consistent with the idea that $\alpha$ variations play
at least some role.  A larger mass density at the onset of an outburst
may lead to a higher $\alpha$ when the instability is triggered.  If
this is the case, then it will naturally follow that both the peak
luminosity of the outburst, and the luminosity at which the
hard-to-soft transition occurs, will be correlated with one another
through the connection to $\alpha$, as has been shown \citep*{yuet04}.

The radiative efficiency of the hot disc will depend on which $L_{\rm NIR}$--$L_{\rm X}$ track
one chooses, and hence on $\alpha$.  In general, the X-ray luminosity
is given by $L_{\rm X}=\epsilon \dot{m}$, where $\epsilon$ is the radiative
efficiency.  For a solution where $L_{\rm X}\propto\dot{m}^2$, then
$\epsilon\propto\dot{m}$.  More specifically, when the two hard state
tracks must join onto the same radiatively efficient soft state track,
then $\epsilon\propto\frac{\dot{m}}{\dot{m_{\rm cr}}}$, where
$\dot{m_{\rm cr}}$ is the mass accretion rate at the critical state transition luminosity (which has one value for the rise track and a different one for the decline). Let us now take a general form for the $\alpha$ dependence of this value, $\dot{m_{\rm cr}}=\alpha^c$.  The radiative
efficiency of the hot disc will then scale as $\alpha^{-c}$.  As a
result, $\dot{m}$ at a given X-ray luminosity will scale as
$\alpha^{-c/2}$, since $L_{\rm X} \propto \dot{m}^2$ in the hard state.

The $\alpha$ variations will thus not lead to parallel tracks in $L_{\rm NIR}$
versus $L_{\rm X}$ if the state transition luminosity scales as
$\alpha^2$.  The $\alpha$ dependence of the jet power extraction
efficiency will cancel out exactly with the $\alpha$ dependence of the
X-ray radiative luminosity.  The estimate of the state transition
luminosity given by Esin et al. (1997) can therefore be valid only if
the fudge factors vary hysteretically or if the radiative efficiency
of the jet varies hysteretically.  The dependence of the transition
luminosity on $\alpha$ must be weaker than $\alpha^2$ in order for
$\alpha$ variations to be solely responsible for the parallel tracks in $L_{\rm X}$
versus $L_{\rm NIR}$.

The estimate of the state transition luminosity put forth by Zdziarski
(1998) does have a weaker than $\alpha^2$ dependence for the state
transition luminosity: $\alpha^{7/5}$.  This leads to a relation where
$L_{\rm X}\propto L_{\rm jet} \alpha^{3/5}$, a closer match to
the observations, but still problematic quantitatively.  For a
given $L_{\rm NIR}$, the X-ray luminosity is $\sim 10$ times higher in the rising
hard state than the decaying one.  The difference
between the X-ray luminosity at the hard-to-soft transition compared to the
soft-to-hard transition is a factor of about five.  If the observations
are to be interpreted in terms of variations in $\alpha$ only, then
either the state transition luminosity must scale slightly weaker than
linearly with $\alpha$, or the jet power must scale more strongly than
linearly with $\alpha$ (e.g. if the fudge factors in the Meier 2001
relation scale with $\alpha$ in some way).

\subsection{Hysteresis traces jet behaviour?}

The other alternative is that the spectral shape or the radiative efficiency of the jet may change
hysteretically.  A test of whether the jet power, or the jet spectrum is the most
important factor may come from correlated high resolution timing in
the X-rays and optical/infrared bands.  It has previously been shown
that a negatively lagged anti-correlation of
optical emission relative to X-ray emission (Kanbach et al. 2001;
Hynes et al. 2003) can be explained as a result of the jet taking away
the bulk of the kinetic luminosity from the accretion flow (Malzac,
Merloni \& Fabian 2004).  If the hysteresis effect we see is the
result of a different amount of kinetic power being taken away by the
jet at the same accretion disc $\dot{m}$, then there should be a signature in the magnitude of the
negatively lagged anti-correlation.  We note that observing the
spectral changes directly would be preferable, but this is quite
difficult due to the convergence around the $I$-band of the optical
components from the accretion disc, the donor star, and the jet in
most systems \citep[e.g.][]{homaet05,russet06}.

A completely independent emission mechanism does exist for explaining
the X-ray emission in X-ray binaries -- jet emission (e.g. Markoff,
Falcke \& Fender 2001; Markoff, Nowak \& Wilms 2005).  In this scenario, the observed parallel tracks would form a jet--jet hysteresis as opposed to a jet--inflow (e.g. jet--corona) hysteresis. In the pure
jet-synchrotron model (Markoff et al. 2001), the NIR-to-X-ray ratio can
change at a given $L_{\rm NIR}$ only if either the spectral slope changes,
or the break frequency where the jet becomes optically thin changes
(see e.g. Nowak et al. 2005 for a discussion of attempts to test the
model based on observed spectral shapes).  In the more recent, more
complicated scenarios, where the corona forms the `base of the jet'
(Markoff et al. 2005) there are additional parameters which can
change this flux ratio, by changing the fraction of the power that
comes from Compton scattering at the base of the jet.  Given that the
physics of the underlying accretion flow feeding the jet is
considerably less developed in these models than for ADAF and related
models, there are no quantitative determinations of what should be
the state transition luminosities or how the resulting hysteresis in the state
transitions should behave.  It thus seems premature to
comment on whether the observations presented in this paper have any
implications for or against the viability of jet X-ray models, but we
note that we here provide an additional constraint for such
models.

\subsection{Consequences of the interpretation}

These findings indicate, on both observational and
theoretical grounds, that there is reason to expect substantial
scatter in relations between radio/NIR and X-ray luminosity especially if
some of the data points are taken during the rising parts of
outbursts.  At the present time, it seems that most of the scatter in
the relations of Gallo et al. (2003, 2006) and \cite{russet06} are due to the
uncertainties in the measurements of the masses and distances of the
different systems used to make the correlations, however two recently discovered sources could be exceptions (\citealt{cadoet07,rodret07}; see also \citealt{gall07}).  As more data is
collected for some of the sources (especially GX 339-4), it may become
clear that the relation is multiple-valued, and the hysteresis effects
will need to be considered.

\section{Conclusions}

We have shown that there is an intrinsic correlation between NIR (which we show comes from the jets) and X-ray luminosity; $L_{\rm NIR}\propto L_{\rm X}^{0.5-0.7}$ for all three low-mass black hole X-ray binaries in the low/hard state which have well sampled data. This is not just a global correlation as shown in \cite{russet06}, but a correlation that holds in all hard state rise and declines. In the high/soft state, there is evidence for the NIR emission to be dominated by the viscously heated disc in all three sources, with $L_{\rm NIR}\propto L_{\rm X}^{0.1-0.2}$ observed in two sources.

The normalisation of the hard state correlation can differ with time for a single source. The infrared flux in the 2000 outburst of XTE J1550--564 shows a hysteresis effect, in that the rising low/hard state
has a lower infrared flux at a given X-ray flux than the decaying low/hard
state.  Several theoretical explanations for this effect seem
plausible, and have direct implications for the accretion process, jet power and the dominating X-ray emission processes.  We have ruled out the possibility that the hysteresis is directly caused by different radiative efficiencies in the inner accretion flow on the rise and decline. The effect may be due to changes in the viscosity parameter $\alpha$ and/or the spectrum/radiative efficiency of the jet. Hopefully, these results will help motivate future
observational searches for more examples of jet-disc hysteresis
effects and more theoretical work which might help to understand
whether the disc viscosity parameter $\alpha$ should vary as a function of time over X-ray binary outbursts.

\section{Acknowledgments}
TJM is grateful to Charles Bailyn, Paolo Coppi, Raj Jain and Mike
Nowak for conversations several years ago, and Chris Done for a
constructive and useful referee's report on an earlier paper, which
helped motivate some of the ideas in this paper.  We are furthermore
grateful to Rob Fender and Emmi Meyer-Hofmeister for more recent
discussions.  Some results are provided by the ASM/RXTE teams at MIT and at the RXTE SOF and GOF at NASA's GSFC. These results are based in part on data taken with the
YALO telescope.  Its queue-mode scheduling was essential for allowing
the production of the well-sampled light curves which are essential
for work of this nature.  In addition to those already cited for their
data analysis work, we are grateful to the YALO consortium members for
supporting this unique and highly useful system of telescope
scheduling, and are grateful to the service observers, data
processors, and queue managers who made the YALO program a success.
EGK acknowledges funding via a Marie Curie Intra-European Fellowship under contract no. MEIF-CT-2006-024668.

\end{document}